\documentclass[12pt]{article}
\usepackage{amsmath}
\usepackage{amssymb}
\date{}
\begin{document}

\title{Closed form Solutions to Some Nonlinear equations
by a Generalized Cole-Hopf Transformation}

\author{Mayer Humi\\
Department of Mathematical Sciences,\\
Worcester Polytechnic Institute,\\
100 Institute Road,\\
Worcester, MA  0l609}

\maketitle

\begin{abstract}
In the first part of this paper we linearize and solve the Van der Pol 
and Lienard equations with some additional nonlinear terms by the 
application of a generalized form of Cole-Hopf transformation. We then show 
that the same transformation can be used to linearize Painleve III equation 
for certain combinations of its parameters. Finally we linearize new forms 
of Burger's and related convective equations with higher order 
nonlinearities. 
\end{abstract}

\thispagestyle{empty}

\newpage

\section{Background}

While there is no existing general theory for integrating nonlinear
ordinary and partial differential equations, the derivation of exact 
(closed form) solutions for these equations is a non-trivial and important 
problem. Some methods to this end are Lie, and related symmetry methods [8]. 
Many other ad-hoc methods for solving nonlinear differential 
equations were independently suggested during the last decades [18].
Among these methods Cole-Hopf transformation [9,10,11,12] has been used 
originally to linearize the Burger's equation [13] 
\begin{equation}
\label{1.1}
\frac{\partial\psi}{\partial t}+\psi\frac{\partial\psi}{\partial x}=
\nu\frac{\partial^2\psi}{\partial x^2},\,\,\,\, \nu=constant.
\end{equation}
This equation contains a convective term and serves as a prototype for 
turbulence modeling, gas dynamics and traffic flow. Such equations 
with convective terms appear in various applications
in applied mathematics and theoretical physics.

Using Cole-Hopf transformation [9,10]
\begin{equation}
\label{1.2}
\psi=\frac{\frac{\partial\phi}{\partial x}}{\phi}.
\end{equation}
it was found that (\ref{1.1}) can be linearized and reduced to the 
linear Heat equation
\begin{equation}
\label{1.3}
\frac{\partial\phi}{\partial t}=k\frac{\partial^2\phi}{\partial x^2}.
\end{equation}
Since this discovery many attempts were made in the literature to 
generalize this result to larger class of equations or to relate other 
nonlinear equations to this equation[11,12].

Another important nonlinear equation is the Van der Pol equation [1]
which without forcing is
\begin{equation}
\label{3.1}
\psi(x)''=\mu(\beta-\psi(x)^2)\psi(x)'-\alpha\psi(x),
\end{equation}
where $\alpha$ and $\beta$ and $\mu$ are arbitrary constants.
This equation was derived originally to model electrical circuits in 
vacuum tubes. However since then it has been used to model phenomena 
in both the physical [2-6] and biological sciences [7]. The solutions of
this equation were investigated extensively (both numerically and analytically) in the literature [2-6,16,17].  

Recently however, we introduced [14] a generalization of Cole-Hopf 
transformation and used it to linearize and solve various nonlinear ordinary 
differential equations e.g Duffing equation. We showed also that many of the 
special functions of mathematical physics are exact solutions for a class 
of nonlinear equations. In this paper we apply in Sec. $2$ this 
generalization of Cole-Hopf transformation to linearize the Van der Pol 
equation with additional quadratic and cubic nonlinear terms. These terms 
can be considered as perturbations to the original equation. Next in Sec. $3$ 
we present explicit solutions (in closed analytic form) to this equation 
and related (polynomial) Lienard equations with and without external forcing.

In Sec $4$ we consider Painleve III equation and show that it
can be linearized by the same transformation for certain combinations
of its parameters.

In Sec. $5$ we consider a class of Burger's equations with additional 
quadratic nonlinear terms which can be solved by the same transformation.
Next in Sec. $6$ we consider second nonlinear ordinary differential 
equations (ODEs) with convective term and derive conditions under which 
they can be linearized. These type of equations represent steady state 
convection in one dimension [15].We end up in Sec. $7$ with a summary.

We wish to point out that we do not include in this paper direct physical 
applications to the solutions of the equations treated in this paper. 
However there exist an extensive research litrature devoted to 
these solutions and their applications. Accordingly the method presented 
in this paper is generic to the class of methods in mathematical physics. 

\setcounter{equation}{0} 
\section{The Perturbed Van der Pol Equation}

In this section we consider Van der Pol equation (\ref{3.1})
with additional nonlinear and forcing terms
\begin{equation}
\label{3.2}
\psi(x)''=\mu(\beta-\psi(x)^2)\psi(x)'-\alpha\psi(x)+v(x)\psi(x)^2+h(x)\psi(x)^3+g(x)\psi(x)^4+f(x).
\end{equation}

We shall say that the solutions of (\ref{3.2}) and
\begin{equation}
\label{3.3a}
\phi(x)''=U(x)\phi(x)
\end{equation}
are related by a generalized Cole-Hopf transformation if we can 
find a function $P(x)$ so that 
\begin{equation}
\label{3.4a}
\psi(x)=P(x)+\frac{\phi(x)^{\prime}}{\phi(x)}.
\end{equation}

To classify those nonlinear equations of the form (\ref{3.2}) whose solutions 
can be obtained from those of the linear equation (\ref{3.3a}) we differentiate 
(\ref{3.4a}) twice and in each step replace the second order derivative of 
$\phi(x)$ using (\ref{3.3a}). Substituting these results in (\ref{3.2}) 
leads to the following equation
\begin{eqnarray}
\label{3.3}
a_4(x)\left(\frac{\phi(x)'}{\phi(x)}\right)^{4}+
a_3(x)\left(\frac{\phi(x)'}{\phi(x)}\right)^{3}+
a_2(x)\left(\frac{\phi(x)'}{\phi(x)}\right)^{2}+
a_1(x)\frac{\phi(x)'}{\phi(x)}+a_0(x)=0
\end{eqnarray}
where
\begin{equation}
\label{3.4}
a_4(x)=-\mu-g(x)
\end{equation}
\begin{equation}
\label{3.5}
a_3(x)=-[2\mu+4g(x)]P(x)-h(x)+2
\end{equation}
\begin{equation}
\label{3.6}
a_2(x)=\mu P(x)'-(6g(x)+\mu)P(x)^2-3h(x)P(x)-v(x)+\mu U(x)+\mu\beta
\end{equation}
\begin{equation}
\label{3.7}
a_1(x)=-4g(x)P(x)^3-3h(x)P(x)^2+[2\mu P(x)'-2v(x)+2\mu U(x)]P(x)-2U(x)+\alpha
\end{equation}
\begin{eqnarray}
\label{3.8}
a_0(x)&=&P(x)^{\prime\prime}+\mu[P(x)^2-\beta)P(x)^{\prime}+U(x)^{\prime}-g(x)P(x)^4 \\ \notag
&&-h(x)P(x)^3+[\mu U(x)- v(x)]P(x)^2+\alpha P(x)-\mu\beta U(x)-f(x)
\end{eqnarray}

To satisfy (\ref{3.3}) it is sufficient to let $a_i(x)=0$, $i=1,2,3,4$.
From $a_4=0$ we obtain
\begin{equation}
\label{3.9}
g(x)=-\mu. 
\end{equation}
Using this result and $a_3(x)=0$ we solve for $h(x)$ (in terms of $P(x)$)
\begin{equation}
\label{3.10}
h(x)=2(\mu P(x)+1). 
\end{equation}
The condition $a_2(x)=0$ can be solved then for $v(x)$
\begin{equation}
\label{3.11}
v(x)=\mu P(x)'+\mu U(x)-[\mu P(x)+6]P(x)+\mu\beta.
\end{equation}
substituting (\ref{3.9})-(\ref{3.11}) in $a_1=0$ we solve for $U(x)$ 
in terms of $P(x)$
\begin{equation}
\label{3.12}
U(x)=3P(x)^2-\mu\beta P(x)+\frac{\alpha}{2}
\end{equation}
Finally using the expressions derived in (\ref{3.9})-(\ref{3.12}) in 
$a_0=0$ we have
\begin{equation}
\label{3.13}
f(x)=P(x)^{\prime\prime}-2\mu\beta P(x)^{\prime}+[6P(x)'+\alpha+\mu^2\beta^2]P(x)+4[P(x)-\mu\beta]P(x)^2-\frac{\mu\beta\alpha}{2}
\end{equation}
This equation can be solved (in principle) for $P(x)$ for a given $f(x)$ 
or it can be used to determine $f(x)$ for apriori choice of $P(x)$. 
Eqs. (\ref{3.10})-(\ref{3.12}) can be solved then (in reverse order) 
to compute the functions $U(x),v(x)$ and $h(x)$.

Using this algorithm we provide in the next section explicit solutions to 
(\ref{3.2}) under various conditions.

\setcounter{equation}{0} 
\section{Solutions to the Van der Pol Equation}

\subsection{Van der Pol Equation with No External Forcing}

When $f(x)=0$ the general solution of (\ref{3.13}) for $P(x)$ is
\begin{equation}
\label{4.1}
P(x)=\frac{2C_1\mu\beta e^{\frac{\mu\beta x}{2}}+C_2(\mu\beta+k)e^{\frac{kx}{2}}+
(\mu\beta-k)e^{-\frac{kx}{2}}}
{4[C_1e^{\frac{\mu\beta x}{2}}+C_2e^{\frac{kx}{2}}+e^{-\frac{kx}{2}}]}
\end{equation}
where $k^2=\mu^2\beta^2-4\alpha$ and $C_1,C_2$ are arbitrary constants.
Similar (but more cumbersome) expressions can be obtained for $P(x)$ when
$f(x)=c$ where $c$ is a non zero constant.

It is now only a matter of simple algebra to compute the general form of
the functions $U(x),v(x)$ and $h(x)$ and then derive solutions to (\ref{3.2})
by solving the linear equation (\ref{3.3a}). 

We consider some simple cases.

{\bf Case} 1: $C_1=C_2=0$.
In this case (\ref{4.1}) reduces to 
\begin{equation}
\label{4.2}
P(x)=\frac{\mu\beta-k}{4}.
\end{equation}
The resulting expressions for $U(x),v(x)$ and $h(x)$ are
\begin{equation}
\label{4.3}
U(x)=\alpha/2+\frac{(3k+\mu\beta)(k-\mu\beta)}{16}
\end{equation}
\begin{equation}
\label{4.4}
v(x)=-\frac{\mu^3\beta^2}{8}+\frac{\mu}{2}\left(\alpha-\beta+\frac{k^2}{4}\right)
+\frac{3k}{2}
\end{equation}
\begin{equation}
\label{4.5}
h(x)=\frac{\mu}{2}(\mu\beta-k)+2.
\end{equation}
With these settings the general solution of (\ref{3.3a}) is 
\begin{equation}
\label{4.5a}
\phi(x)=C_3\cos(\omega x)+C_4\sin(\omega x)
\end{equation}
where
$\omega=\frac{1}{4}\sqrt{\mu^2\beta^2+2\mu\beta k-3k^2-8\alpha}$.
These solutions are related to the solutions of (\ref{3.2}) by
the transformation (\ref{3.4a}). 

{\bf Case} 2: Assume $k=0$ and $C_1=0$.

In this case 
\begin{equation}
\label{4.6}
P(x)=\frac{\mu\beta}{4},\,\,\, U(x)=\frac{\alpha}{2}-\frac{\mu^2\beta^2}{16}
,\,\,\,\, h(x)=\frac{\mu^2\beta}{2}+2,\,\,\, 
v(x)=\frac{\mu}{2}\left(\alpha-\beta-\frac{\mu^2\beta^2}{4}\right)
\end{equation}
The solution for $\phi(x)$ is in the same form as in (\ref{4.5a}) with
$\omega=\frac{1}{4}\sqrt{\mu^2\beta^2-8\alpha}$.

{\bf Case} 3: Assume $\alpha=0$.

When $\alpha=0$, $k=\pm \mu\beta$. (In the following we consider only
the plus sign). Under this constraint (\ref{4.1}) reduces to
\begin{equation}
\label{4.7}
P(x)=\frac{(C_1+C_2)\mu\beta e^{\frac{\mu\beta x}{2}}}
{(C_1+C_2+e^{-\mu\beta x}}
\end{equation}
Using (\ref{3.12}) yields
\begin{equation}
\label{4.8}
U(x)=-\frac{c\mu^2\beta^2(e^{-\mu\beta x}-c)}{e^{-\mu\beta x}+c)^2}
\end{equation}
where $c=C_1+C_2$
Similarly (\ref{3.11}) and (\ref{3.12}) lead to
\begin{equation}
\label{4.9}
v(x)=\frac{\mu\beta(e^{-\mu\beta x}-2c)}{e^{-\mu\beta x}+c},\,\,\,\,
h(x)=2+\frac{\mu^2\beta c}{e^{-\mu\beta x}+c}
\end{equation}
With this data the general solution for (\ref{3.3a}) is
\begin{equation}
\label{4.10}
\phi(x)=\frac{C_3+C_4(ce^{\mu\beta x}+\mu\beta x)}{\sqrt{1+ce^{\mu\beta x}}}
\end{equation}

\subsection{Van der Pol Equation with External forcing}

When the forcing function $f(x)$ is not zero or a constant (\ref{3.13})
can not be solved in closed form for $P(x)$. It is expedient in this 
case to reverse the process and classify those equations of the form 
(\ref{3.2}) which can be linearized for a given $P(x)$.

To make a proper choice of $P(x)$ we observe that in order to obtain
closed form solutions for $\psi(x)$ the function $U(x)$ has to be in a form
which enables the solution of (\ref{3.3a}) to be expressed in "simple form".

To carry this out it is easy to see from (\ref{3.12}) that when 
$$
U(x)= 3g(x)^2-\mu\beta g(x)+\frac{\alpha}{2},
$$
(where g(x) is an arbitrary smooth function) then
$$
P(x)= g(x),\,\,\, or \,\,\, P(x)=-g(x)+\frac{\mu\beta}{3}
$$
The computation of the functions $f(x)$, $v(x)$ and $h(x)$ can be carried out 
then by substitution. We present some examples.

{\bf Example} 1: Let $g(x)=ax$ where a is a constant. We then have
$$
U(x)=3a^2x^2+a\mu\beta x-\frac{\alpha}{2},
$$
$$
f(x)=-4a^3x^3+a\left(6a-\alpha+\frac{\mu^2\beta^2}{3}\right)x-
\frac{\mu\beta\alpha}{6}+\frac{\mu^3\beta^3}{27}
$$
The general solution for $\phi(x)$ can be expressed then in terms of
Hypergeometric functions. 

{\bf Example} 2: let $g(x)=tan(x)$. In this case we obtain an oscillatory 
forcing function and the solutions of (\ref{3.3a}) for $\phi(x)$ can be 
expressed again in terms of Hypegeometric functions.

\subsection{Polynomial Lienard Equations}

The general form of Lienard equations is
\begin{equation}
\label{5.1}
\frac{d^2\psi}{dt^2}+f(\psi)\frac{d\psi}{dt}+g(\psi)=0 
\end{equation}
where $f(\psi)$ and $g(\psi)$ are smooth functions.

Since Van der Pol equation with no external forcing belongs to this
class of equations it is appropriate to ask if the same method used to solve
(\ref{3.2}) can be applied to this more general class of equations. In this 
section we explore this application to (\ref{5.1}) when $f(x)$ and $g(x)$ are 
respectively a second and fourth order polynomials
$$
f(x)=\displaystyle\sum_{k=0}^2 c_k\psi(x)^k,\,\,\,\,
g(\psi)=\displaystyle\sum_{k=0}^4 b_k\psi(x)^k
$$  
Applying the transformation (\ref{3.4a}) with $\phi(x)$ satisfying 
(\ref{3.3a}) to (\ref{5.1}) leads to an equation similar to (\ref{3.3}).
which can be satisfied if we set $a_i=0$ for $i=0,\ldots 4$.
We solve this set of equations for $b_i$ in terms of $c_i$, $P(x)$
and $U(x)$. We obtain the following relations:
\begin{equation}
\label{5.2}
b_4(x)=c_2(x),\,\,\,\, b_3(x)=c_1(x)-2c_2(x)P(x)+2,\,\,\,\,
\end{equation}
\begin{equation}
\label{5.3}
b_2(x)=c_2(x)P(x)^2-2(3-c_1(x))P(x)-c_2(x)(P(x)'-U(x))+c_0(x)
\end{equation}
\begin{equation}
\label{5.4}
b_1(x)=(6+c_1(x))P(x)^2-2c_0(x)P(x)-c_1(x)P(x)'-(c_1(x)+2)U(x),\,\,\,\,
\end{equation}
\begin{equation}
\label{5.5}
b_0(x)=P(x)''-c_0(x)P(x)'+U(x)'-2P(x)^3+2P(x)U(x)+c_0(x)(P(x)^2-U(x))
\end{equation}

For apriori choice of the functions $c_i(x)$ and $U(x),P(x)$ these 
equations can be solved for the $b_i(x)$. When these relations hold 
we obtain an equation of the form (\ref{5.1})
whose solutions are related to those of (\ref{3.3a}) by the 
transformation (\ref{3.4}).

An interesting case arises when $b_0=0$. Under this condition (\ref{5.5})
will be satisfied if
\begin{equation}
\label{5.6}
U(x)=P(x)^2-P(x)'
\end{equation}
which is a Riccati equation for $P(x)$ and $c_0(x)$ remains a free parameter.

\setcounter{equation}{0}
\section{Painleve III Equation}

The general form of Painleve III equation is
\begin{equation}
\label{6.1}
\psi^{\prime\prime}=\frac{(\psi^{\prime})^2}{\psi}-\frac{\psi^{\prime}}{x}+
\frac{\alpha \psi^2+\beta}{x}+\gamma \psi^3+\frac{\delta}{\psi}
\end{equation}
where $\alpha,\,\beta\,\,\gamma,\,\delta$ are parameters.

We shall say that we can linearize this equation (for certain values 
of its parameters) if we can find functions $P(x)$,$Q(x)$ so that the 
functions $\psi(x)$, $\phi(x)$ are related by eq. (\ref{3.4a}) and 
$\phi(x)$ satisfies the linear equation
\begin{equation}
\label{6.2} 
\phi(x)''=K(x)\phi(x)'+U(x)\phi(x).
\end{equation}
Following the same steps as in Sec $2$ we obtain an equation of the form
(\ref{3.3}) where
\begin{equation}
\label{6.3}
a_4(x)=Q(x)^2-\gamma Q(x)^4
\end{equation}
\begin{equation}
\label{6.4}
a_3(x)=\frac{Q(x)}{x}[2P(x)x+\alpha Q(x)^2\alpha+Q(x)K(x)x+Q(x)]
\end{equation}
\begin{eqnarray}
\label{6.5}
&&a_2(x)=\frac{1}{x}[2xQ(x)P(x)'+Q(x)Q(x)'-3\alpha P(x)Q(x)^2- \\ \notag
&&x(Q(x)')^2-3P(x)Q(x)K(x)-P(x)Q(x)+xQ(x)Q(x)''+xQ(x)^2K(x)'- \\ \notag
&&6xP(x)^2+Q(x)^2K(x)-2xP(x)Q(x)']
\end{eqnarray}
\begin{eqnarray}
\label{6.6}
&&a_1(x)=\frac{1}{xQ(x)}\{[-U(x)-xU(x)'+xK(x)U(x)]Q(x)^3+[-K(x)P(x)- \\ \notag
&&xP(x)K(x)'-xP(x)K(x)^2+ 2xP(x)'K(x)+\beta-P(x)'-xP(x)''+3\alpha P(x)^2+ \\ \notag
&&2xP(x)U(x)]Q(x)^2+[-xP(x)Q(x)''-P(x)Q(x)'+2xP(x)'Q(x)' \\ \notag
&&-2xP(x)K(x)Q(x)']Q(x)+4xP(x)^3\}
\end{eqnarray}
\begin{eqnarray}
\label{6.7}
&&a_0(x)=-\frac{P(x)^3(xP(x)+\alpha Q(x)^2)}{xQ(x)^2}+ \\ \notag
&&\{P(x)''-\frac{\beta}{x}+[K(x)U(x)+U(x)'+\frac{U(x)}{x}]Q(x)+ \\ \notag
&&2Q(x)'U(x)+\frac{P(x)'}{x}\}P(x)-Q(x)^2U(x)^2-2P(x)'Q(x)U(x)-(P(x)')^2-\delta
\end{eqnarray}
To satisfy (\ref{3.3}) it is sufficient to let $a_i(x) = 0$, 
$i = 1,\, 2,\, 3,\, 4$. From $a_4 = 0$ it follows that $Q(x)^{-2}=\gamma$
i.e Q(x) a constant. (Hence, in the following, we refer to it as $Q$). 
Using this result and solving (\ref{6.4}) for $K(x)$  we have
\begin{equation}
\label{6.8}
K(x)=-\frac{2xP(x)+\alpha Q^2+Q}{xQ}
\end{equation}
Inserting these results in (\ref{6.5}) we find that it being satisfied
automatically (no further restrictions). Eq (\ref{6.6}) reduces then 
to a first order linear equation for $U(x)$ in terms of $P(x)$ whose general 
solution is
\begin{equation}
\label{6.9}
U(x)=-\frac{P(x)'}{Q}-\frac{P(x)^2}{Q^2}-\frac{\alpha Q+1}{xQ}P(x)+
\frac{\beta}{Q(\alpha Q+2)}+Cx^{-(\alpha Q+2)}
\end{equation}
where $C$ is an arbitrary constant. Substituting all these results
in (\ref{6.7}) we find that it reduces to
\begin{equation}
\label{6.10}
\delta+\frac{\beta^2}{(\alpha Q+2)^2}+CQ[\frac{2\beta}
{(\alpha Q+2)x^{\alpha Q+2}}+\frac{CQ}{x^{2\alpha Q+4}}]=0
\end{equation}
(viz. the coefficient of $P(x)$ is zero and therefore $P(x)$ remains
as a free parameter function). Eq. (\ref{6.10}) is satisfied if we let
$C=0$ and 
\begin{equation}
\label{6.11}
\delta=-\frac{\beta^2}{(\alpha Q+2)^2}
\end{equation}
When (\ref{6.11}) is satisfied Painleve III equation can be linearized.
We can summarize these results by the following lemma:

{\bf Lemma} Let the parameters in eq. (\ref{6.1}) satisfy eq. (\ref{6.11})
(with $Q^2=1/\gamma$). For any (smooth) choice of the function $P(x)$ the 
solutions of eq. (\ref{6.2}) where $K(x)$, $U(x)$ are given by eqs. (\ref{6.8}),
(\ref{6.9}), provide solutions to eq. (\ref{6.1}) through eq. (\ref{3.4a}). 

{\bf Example 1}: If we let the parameters in eq. (\ref{6.1}) be chosen as
$$
\alpha=-1,\,\,\beta=1,\,\, \gamma=1\,\,(Q=1)\,\, and\,\, \delta=-1
$$
then eq. (\ref{6.11}) is satisfied. For 
$$
P(x)= ax+bx^2+cx^3+d
$$
we have
$$
K(x)=-2P(x),\,\,\, U(x)=-P(x)^2-P(x)'+1.
$$
The general solution for $\phi$ is
$$
\phi(x) =C_1\exp\{-\frac{x}{2}[\frac{cx^3}{2}+\frac{2bx^2}{3}+ax+2d-2]\}+
C_2\exp\{-\frac{x}{2}[\frac{cx^3}{2}+\frac{2bx^2}{3}+ax+2d+2]\}
$$
The solution $\psi(x)$ to eq. (\ref{6.1}) is given then by eq. (\ref{3.3a})
for any values of the integration constants $C_1,\,C_2$ and this can 
verified by direct substitution. (This demonstrates also that the 
superposition principle holds for these solutions).

{\bf Example 2}: With the same parameters for eq. (\ref{6.1}) as in example 
$1$ we let 
$$
P(x)=\sin x
$$
Computing the corresponding expressions for $K(x)$ and $U(x)$ we find that
the solution of eq. (\ref{6.2}) is
$$
\phi(x)=C_1e^{\cos x}\sinh x + C_2e^{\cos x}\cosh x.
$$
Once again the solution $\psi(x)$ to eq. (\ref{6.1}) is given then by 
eq. (\ref{3.3a}).

{\bf Example 3}:With the same parameters for eq. (\ref{6.1}) as in example
$1$ we let
$$
P(x)=xe^x
$$
The solution for $\phi(x)$ is
$$
\phi(x)=C_1\exp((1-x)e^x)\sinh x +C_2\exp((1-x)e^x)\cosh x
$$
and $\psi(x)$ is given by (\ref{3.3a}).

\setcounter{equation}{0} 
\section{Generalized Burger Equation}
In this section we consider equations of the form
\begin{equation}
\label{14.1}
\frac{\partial\psi}{\partial t}-M(x)\frac{\partial^2\psi}{\partial x^2} =
H(x)\psi\frac{\partial\psi}{\partial x}+V(x)\psi+W(x)\psi^2
\end{equation} 
where $\psi=\psi(x,t)$.
In this case we shall say that the solutions of this equation are related
to the solution of the linear equation
\begin{equation}
\label{14.2}
\frac{\partial\phi}{\partial t}-M(x)\frac{\partial^2\phi}{\partial x^2}=0
\end{equation}
if we can find functions $P(x)$, $Q(x)$ so that
\begin{equation}
\label{14.3}
\psi=P(x)+Q(x)\frac{\frac{\partial\phi}{\partial x}}{\phi}
\end{equation}

To classify those equations of the form (\ref{14.1}) which can be paired 
to the linear equation (\ref{14.2}) we substitute (\ref{14.3}) in (\ref{14.1}).
After some algebra we find that the following equation has to be satisfied
\begin{equation}
\label{14.4}
a_3\left(\frac{\frac{\partial\phi}{\partial x}}{\phi}\right)^{3}+
a_2\left(\frac{\frac{\partial\phi}{\partial x}}{\phi}\right)^{2}+
a_{21}\left(\frac{\frac{\partial\phi}{\partial x}}{\phi^2}\right)+
a_1\frac{\frac{\partial\phi}{\partial x}}{\phi}+
\frac{a_{11}}{\phi(x)}+
a_0=0,
\end{equation}
where
\begin{equation}
\label{14.5}
a_{21}=-Q(x)\left[(Q(x)H(x)-3M(x))\frac{\partial^2\phi}{\partial x^2}+
\frac{\partial\phi}{\partial t}\right],
\end{equation}
\begin{equation}
\label{14.6}
a_{11}=-(Q(x)H(x)P(x)+2M(x)Q(x)')\frac{\partial^2\phi}{\partial x^2}+
Q(x)\frac{\partial^2\phi}{\partial t\partial x} -
M(x)Q(x)\frac{\partial^3\phi}{\partial x^3},
\end{equation}
\begin{equation}
\label{14.7}
a_{3}=2M(x)-Q(x)H(x),
\end{equation}
\begin{equation}
\label{14.8}
a_{2}=H(x)Q(x)P(x)+(2M(x)-H(x)Q(x))Q(x)'-W(x)Q(x)^2,
\end{equation}
\begin{equation}
\label{14.9}
a_1=-H(x)Q(x)P(x)'-H(x)P(x)Q(x)'-M(x)Q(x)''-V(x)Q(x)-2W(x)P(x)Q(x),
\end{equation}
\begin{equation}
\label{14.10}
a_{0}=-M(x)P(x)'' - H(x)P(x)P(x)'-W(x)P(x)^2-V(x)P(x),
\end{equation}
where primes denote differentiation with respect to $x$.
To satisfy (\ref{14.4}) it is sufficient to let $a_{21}=a_{11}=0$
and $a_3=a_2=a_1=a_0=0$.

From $a_3=0$ it follows that
\begin{equation}
\label{14.11}
Q(x)=\frac{2M(x)}{H(x)}
\end{equation}
In view of this relationship $a_{21}=0$ is satisfied in virtue of (\ref{14.2}).
To satisfy $a_{11}=0$ we add and subtract 
$M(x)'\frac{\partial^2\phi}{\partial x^2}$ and rewrite this equation in 
the form
\begin{equation}
\label{14.12}
a_{11}=-(Q(x)H(x)P(x)+2M(x)Q(x)'-Q(x)M(x)')\frac{\partial^2\phi}{\partial x^2}+
Q(x)\frac{\partial^2\phi}{\partial t\partial x} -
Q(x)\frac{\partial}{\partial x}\left( M \frac{\partial^2\phi}{\partial x^2}\right)=0.
\end{equation}
Letting 
\begin{equation}
\label{14.13}
P(x)=\frac{2M(x)H(x)'}{H(x)^2}-\frac{M(x)'}{H(x)}
\end{equation}
we infer that (\ref{14.12}) is satisfied using (\ref{14.2}).

We now use $a_2=a_1=0$ to express $W(x)$ and $V(x)$ in terms of $M(x),\,H(x)$.
Substituting the expressions for Q(x) and P(x) in $a_2=0$ we obtain 
\begin{equation}
\label{14.14}
W(x)=-\frac{H(x)M(x)'}{2M(x)}+H(x)'
\end{equation}
Similarly $a_1=0$ yields
\begin{equation}
\label{14.15}
V(x)=-\frac{M(x)H(x)''}{H(x)}
\end{equation}
With these results $a_0=0$ yield a differential equation relating
$M(x)$ and $H(x)$
\begin{equation}
\label{14.16}
\left[\frac{H(x)M(x)'}{2M(x)}-H(x)'\right]P(x)^2+
\left[\frac{M(x)H(x)''}{H(x)}-H(x)P(x)'\right]P(x)
-M(x)P(x)''=0
\end{equation}
where for brevity we did not substitute for $P(x)$.

When $M(x)=A$ where $A$ is a constant (\ref{14.16}) becomes
\begin{equation}
\label{14.17}
H(x)^2H(x)'''-5H(x)H(x)'H(x)''+4(H(x)')^3=0.
\end{equation}
To solve this equation we introduce $z(x)=\frac{1}{H(x)}$. The resulting 
equation can be written as
\begin{equation}
\label{14.18}
\frac{d}{dx}\left(\frac{1}{z(x)}\frac{d^2z(x)}{dx^2}\right)=0.
\end{equation}
Hence either 
\begin{equation}
\label{14.18a}
H(x)=\frac{1}{ax+b}
\end{equation}
where $a,b$ are constants or
\begin{equation}
\label{14.19}
H(x)=\frac{B}{\cos(\omega x + \beta)},\,\,\,\, B,\omega,\beta,\,\,\, constants,
\end{equation}
or
\begin{equation}
\label{14.20}
H(x)=Ce^{\alpha x},\,\,\,\, C,\alpha,\,\,\, constants,
\end{equation}

When $H(x)=1$ eq. (\ref{14.16}) becomes
\begin{equation}
\label{14.21}
\frac{M(x)'}{2M(x)}-M(x)'M(x)''+M(x)M(x)'''=0.
\end{equation}
Substituting $M(x)=w(x)^2$ this equation reduces to
\begin{equation}
\label{14.22}
\frac{d}{dx}\left(w(x)\frac{d^2w}{dx^2}\right) =0.
\end{equation}
Hence
$$
w(x)\frac{d^2w}{dx^2} =c
$$
where $c$ is a constant. When $c=0$ it follows that 
\begin{equation}
\label{14.23}
M(x)=(a_1x+b_1)^2
\end{equation}
where $a_1,b_1$ are constants. When $c \ne 0$ we can find an implicit 
expression $w(x)$
$$
x=C_2\pm \displaystyle\int^{w(x)}\frac{ds}{\sqrt{2c\ln(s)-C_1c}},
$$
where $C_1,\,C_2$ are integration constants.

To our best knowledge the linearization of the "modified Burger's equations" 
represented by (\ref{14.18a})-(\ref{14.20}) and (\ref{14.23}) did not
appear in the literature so far.

{\bf Example}: For $M(x)=1$, $H(x)=Ce^{\alpha x}$ we obtain from (\ref{14.14})
(\ref{14.15}) respectively that
$$
W(x)=C\alpha e^{\alpha x},\,\,\,\, V(x)=-\alpha^2.
$$
Hence the solutions of (\ref{14.1}) with these coefficients is related to the 
solutions of the Heat equation (\ref{14.2}) by the transformation
(\ref{14.3}) with 
$$
Q(x)=\frac{2e^{-\alpha x}}{C},\,\,\,\, P(x)=\frac{2\alpha e^{-\alpha x}}{C}
$$
and this fact can be verified by direct substitution. 

\setcounter{equation}{0} 
\section{Second Order Convective ODEs}

We shall say that the solutions of the equations
\begin{equation}
\label{2.1}
\psi(x)''=S(x)+ [V(x)+F(x)\psi(x)']\psi(x)+W(x)\psi(x)^2
\end{equation}
and
\begin{equation}
\label{2.2}
\phi(x)''=U(x)\phi(x)
\end{equation}
are related if we can find functions $P(x)$ and $Q(x)$ so that
\begin{equation}
\label{2.3}
\psi(x)=P(x)+Q(x)\frac{\phi(x)^{\prime}}{\phi(x)}.
\end{equation}
Furthermore we observe that (\ref{2.1}) can take the more general form
\begin{equation}
\label{2.1b}
\psi(x)'' = S(x)+ (V(x)+F(x)\psi(x)')\psi(x)+V_1(x)\psi(x)'+W(x)\psi(x)^2.
\end{equation}
In this case we can find $p(x)$ so that $V_1(x)=-2\frac{p(x)'}{p(x)}$.
Introducing $\xi(x)=p(x)\psi(x)$, (\ref{2.1b}) becomes
\begin{equation}
\label{2.1c}
\xi(x)''= p(x)S(x)+\left[V(x)+\frac{p(x)''}{p(x)}+\frac{F(x)}{p(x)}\xi(x)'\right]\xi(x)+
\left[\frac{W(x)}{p(x)}-\frac{F(x)p(x)'}{p(x)^2}\right]\xi(x)^2.
\end{equation}
which has the same form as (\ref{2.1}).

To classify those nonlinear equations (\ref{2.1}) which can be "paired"
with a linear equation of the form (\ref{2.2}) we 
follow the same steps as in the previous section and find that the following
equation must hold;
\begin{equation}
\label{2.4}
a_3(x)\left(\frac{\phi(x)'}{\phi(x)}\right)^{3}+
a_2(x)\left(\frac{\phi(x)'}{\phi(x)}\right)^{2}+
a_1(x)\frac{\phi(x)'}{\phi(x)}+a_0(x)=0
\end{equation}
where 
\begin{equation}
\label{2.5}
a_3(x)=(Q(x)F(x)+2),\,\,\,a_2(x)=Q(x)(F(x)P(x)-W(x)Q(x))-(2+F(x)Q(x))Q(x)',
\end{equation}
\begin{eqnarray}
\label{2.6}
a_1(x)&=&Q(x)''-F(x)P(x)Q(x)'-U(x)F(x)Q(x)^2 - \\ \notag
&&(2U(x)+V(x)+F(x)P(x)'+2W(x)P(x))Q(x),
\end{eqnarray}
\begin{eqnarray}
\label{2.7}
a_0(x)&=&P(x)''-W(x)P(x)^2-(F(x)Q(x)U(x)+V(x)+F(x)P(x)')P(x)+ \\ \notag
&&Q(x)U(x)'+2U(x)Q(x)'-S(x).
\end{eqnarray}
To satisfy (\ref{2.4}) it is therefore sufficient to let 
$a_i(x)=0,\,\,i=0,1,2,3$.

As in previous sections one can use these conditions in two ways . 
The first is to assume that 
a nonlinear equation (\ref{2.1}) is given and try to determine the appropriate 
$P,Q,U$ (if they exist) that relates it to (\ref{2.3}). Otherwise one may 
fix the functions $P,Q,U$ and classify those nonlinear equations of the form 
(\ref{2.1}) which are related to (\ref{2.3}) by the transformation (\ref{2.2}).
In the following we provide separate solutions for these two possibilities.

Assuming that one starts from (\ref{2.1}) i.e the functions $V(x),W(x),F(x)$
and $S(x)$ (with $F(x) \ne 0$) are given it follows from (\ref{2.4}) that
\begin{equation}
\label{2.8}
Q(x)=-\frac{2}{F(x)}.
\end{equation}
Substituting this result in (\ref{2.5}) and solving for $P(x)$ it follows that
\begin{equation}
\label{2.9}
P(x)=-\frac{2W(x)}{F(x)^2}.
\end{equation}
Using (\ref{2.8}),(\ref{2.9}) and (\ref{2.7}) we obtain a linear first order 
differential equation for $U(x)$
\begin{eqnarray}
\label{2.10}
&&U(x)'+\frac{F(x)S(x)}{2}+\frac{W(x)''-V(x)W(x)+(2W(x)-2F(x)')U(x)}{F(x)}\\ \notag
&&\frac{(2W(x)-4F(x)')W(x)'-2W(x)F(x)''}{F(x)^2}+
\frac{2W(x)(W(x)^2+3(F(x)')^2-2W(x)F(x)'}{F(x)^3}=0
\end{eqnarray}
Finally eq. (\ref{2.6}) provides (after using (\ref{2.8}) and (\ref{2.9}))
an intrinsic constraint on the functions $V(x),W(x),F(x)$and $S(x)$ which 
have to be satisfied for the relationship between (\ref{2.1}) and (\ref{2.3})
to exist.
\begin{equation}
\label{2.12}
V(x)+\frac{F(x)''-2W(x)'}{F(x)}+\frac{6W(x)F(x)'-2(F(x)')^2-4W(x)^2}{F(x)^2}=0.
\end{equation}

We observe that when $F(x)=0$ the algorithm can be implemented in
the same way by adding a term $R(x)\psi(x)^3$ to eq. (\ref{2.1}). 

For the reverse procedure where one elects the functions $U(x),P(x)$ and 
$Q(x)$ and attempts to evaluate the corresponding nonlinear equation 
(\ref{2.1}). We have from (\ref{2.7})
\begin{equation}
\label{2.13}
F(x)=-\frac{2}{Q(x)},\,\,\, W(x)=-\frac{2P(x)}{Q(x)^2}.
\end{equation}
Substituting these results in (\ref{2.7}) and solving for V(x) yields
\begin{equation}
\label{2.14}
V(x)=\frac{Q(x)Q(x)''+4P(x)^2+2(P(x)Q(x))'}{Q(x)^2}.
\end{equation}
Finally from (\ref{2.8}) we derive  an expression for $S(x)$
\begin{equation}
\label{2.15}
S(x)=P(x)''+Q(x)U(x)'+2(Q(x)'+P(x))U(x)-\frac{P(x)Q(x)''}Q(x)
-\frac{2P(x)^2(Q(x)'+P(x))}{Q(x)^2}.
\end{equation}

{\bf Example}: In the differential equation
\begin{equation}
\label{2.16}
\psi(x)''= [4a^2+\psi(x)']\psi(x)+a\psi(x)^2,
\end{equation}
where  $a$ is a constant we have $F(x)=1$, $W(x)=a$, $V(x)=4a^2$ and 
$S(x)=0$. The equation satisfies the constraint (\ref{2.12}) and using
(\ref{2.8})-(\ref{2.10}) we find that
$$
Q(x)=-2,\,\,\, P(x)=-2a.
$$
From (\ref{2.10}) we find that the general solution for $U(x)$ is
$$
U(x)=Ce^{-2ax}+a^2.
$$
With this $U(x)$ the general solution of (\ref{2.2}) is 
$$
\phi(x)=C_1BesselI(1,z)+C_2BesselY(1,iz)
$$ 
where $z=\sqrt{C}\frac{e^{-ax}}{a}$ and $BesselI$, $BesselY$
are the modified Bessel functions of the first and second kind.
Letting $C_2=0$ (real solution) it is straightforward to verify that
$$
\psi(x)=-2a-2\frac{\phi(x)'}{\phi(x)}
$$
is a solution of (\ref{2.16}).

\section{Summary}

We demonstrated in this paper that a generalized form of Cole-Hopf 
transformation (\ref{3.4}) can be used to find solutions of the perturbed
Van der Pol equation without forcing and for Painleve III equation. 
In addition we applied this 
transformation to linearize a generalized form of Burger's equation and 
second order nonlinear ODEs with convective terms.

\section*{References}

\begin{itemize}

\item[1] B. Van der Pol- A theory of the amplitude of free and forced triode 
vibrations, Radio Rev. 1 (1920) pp 701-710.

\item[2] J. Guckenheimer- Dynamics of the Van der Pol Equation,
IEEE Trans. Circuits and Systems, Vol cas-27 pp.983-989 (1980).

\item[3] J. Guckenheimer and P. Holmes-Nonlinear Oscillations, 
Dynamical Systems,and Bifurcations of Vector Fields, 
Springer-Verlag, New York 1990.

\item[4] G.M. Moremedi, D.P. Mason, V.M. Gorringe- On the limit cycle of a 
generalized van der Pol equation, International Journal of 
Non-Linear Mechanics, Volume 28, Issue 2, March 1993, Pages 237-25.

\item[5]J. Guckenheimer, K. Hoffman, and W. Weckesser -
The Forced van der Pol Equation I: The Slow Flow and Its Bifurcations
SIAM Journal on Applied Dynamical Systems 2003 Vol 2, No. 1, pp. 1-3.

\item[6] Raimond A. Struble and John E. Fletcher- General Perturbational 
Solution of the Harmonically Forced van der Pol Equation,
J. Math. Phys. 2, 880 (1961).

\item[7] FitzHugh R. (1955) Mathematical models of threshold phenomena in the 
nerve membrane. Bull. Math. Biophysics, 17 pp. 257-278.

\item[8] C. Gu, H. Chaohao and Z. Zhou - Darboux Transformations in 
Integrable Systems, Springer, New-York (2005).

\item[9] E. Hopf - The partial differential equation $u_t+uu_x=u_{xx}$,
Commun. Pure Appl. Math {\bf 3}, pp.201-230 (1950).

\item[10] J.D. Cole, On a quasi-linear parabolic equation occurring in 
aerodynamics, Quart. Appl. Math. {\bf 9}, pp.225-236 (1951).

\item[11] P. L. Sachdev-A generalized Cole-Hopf transformation for 
nonlinear parabolic and hyperbolic equations,
ZAMP {\bf 29}, No 6 (1978), pp. 963-970.

\item[12] B. Gaffet-On the integration of the self-similar equations and
the meaning of the Cole-Hopf Transformation J. Math. Phys. 27, 2461 (1986).

\item[13] Burgers J.M. A mathematical model illustrating the theory of 
turbulence, Adv. Appl. Mech., 1948. vol. 1, pp.171-199.

\item[14] M. Humi -A Generalized Cole-Hopf Transformation
for Nonlinear ODES, J. Phys. A: Math. Theor. 46 (2013) 325202 (14pp).

\item [15] A. Bouhamidi, M. Hached and K. Jbilou -A meshless method for the
numerical computation of the solution of steady Burgers-type equations,
Applied Numerical Mathematics, 74, pp.95-110 (2013).

\item [16] A. Buonomo -The periodic solution of van der Pol's equation, 
SIAM J. App. Math, 59, pp. 156-171 (1998).

\item [17] Yuan-Ju Huang, Hsuan-Ku Liu- A new modification of the variational 
iteration method for van der Pol equations, Applied Mathematical Modelling,
37 pp.8118-8130 (2013).

\item[18] C. Gu, H. Hu and Z. Zhou -Darboux Transformations in Integrable Systems,\ Springer,NY (2005).

\item[19]  K. Iwasaki, H Kimura, S Shimomura and M. Yoshida -From Gauss to Painlevé ,\,Springer, NY 1991.

\end{itemize}

\end{document}